\documentclass[10pt]{article}

\usepackage{fullpage}
\usepackage{setspace}
\usepackage{parskip}
\usepackage{titlesec}
\usepackage[section]{placeins}
\usepackage{xcolor}
\usepackage{breakcites}
\usepackage{lineno}
\usepackage{hyphenat}
\usepackage{makecell}
\usepackage{xcolor}
\newcommand{\tg}[1]{\textcolor{gray}{#1}}
\usepackage[margin=1in]{geometry}

\PassOptionsToPackage{hyphens}{url}
\usepackage[colorlinks = true,
            linkcolor = blue,
            urlcolor  = blue,
            citecolor = blue,
            anchorcolor = blue]{hyperref}
\usepackage{etoolbox}

\usepackage[numbers]{natbib}
\usepackage{etoolbox}

\newcommand\EatDot[1]{}
\newcommand\Addstar[1]{*}

\renewenvironment{abstract}
  {{\bfseries\noindent{\abstractname}\par\nobreak}\footnotesize}
  {\bigskip}

\titlespacing{\section}{0pt}{*3}{*1}
\titlespacing{\subsection}{0pt}{*2}{*0.5}
\titlespacing{\subsubsection}{0pt}{*1.5}{0pt}

\usepackage{authblk}

\usepackage{graphicx}
\usepackage[space]{grffile}
\usepackage{latexsym}
\usepackage{textcomp}
\usepackage{longtable}
\usepackage{tabulary}
\usepackage{booktabs,array,multirow}
\usepackage{amsfonts,amsmath,amssymb}
\newif\iflatexml\latexmlfalse

\AtBeginDocument{\DeclareGraphicsExtensions{.pdf,.PDF,.eps,.EPS,.png,.PNG,.tif,.TIF,.jpg,.JPG,.jpeg,.JPEG}}

\usepackage[utf8]{inputenc}
\usepackage[english]{babel}

\usepackage{float}

\begin{document}

\title{Interplay of damage and repair in the control of epithelial tissue integrity in response to cyclic loading}

\author[1]{Eleni Papafilippou}%
\author[2]{Lucia Baldauf}%
\author[2,3,4]{Guillaume Charras}%
\author[1]{Alexandre J. Kabla}%
\author[5]{Aessandra Bonfanti}
\affil[1]{Department of Engineering, University of Cambridge, Cambridge, UK}%
\affil[2]{London Centre for Nanotechnology, University College London, London, UK}%
\affil[3]{Institute for the Physics of Living Systems, University College London, London, UK}%
\affil[4]{Department of Cell and Developmental Biology, University College London, London, UK}%
\affil[5]{Department of Civil and Environmental Engineering, Politecnico di Milano, Milan, Italy}%

\date{}

\vspace{-1em}

\begingroup
\let\center\flushleft
\let\endcenter\endflushleft
\maketitle
\endgroup


\selectlanguage{english}
\begin{abstract}
Epithelial tissues are continuously exposed to cyclic stretch \textit{in vivo}. Physiological stretching has been found to regulate soft tissue function at the molecular, cellular, and tissue scales, allowing tissues to preserve their homeostasis and adapt to challenges. In contrast, dysregulated or pathological stretching can induce damage and tissue fragilisation. Many mechanisms have been described for the repair of epithelial tissues across a range of time-scales. In this review, we present the timescales of (i) physiological cyclic loading regimes, (ii) strain-regulated remodelling and damage accumulation, and (iii) repair mechanisms in epithelial tissues. We discuss how fatigue in biological tissues differs from synthetic materials, in that damage can be partially or fully reversed by repair mechanisms acting on timescales shorter than cyclic loading. We highlight that timescales are critical to understanding the interplay between damage and repair in tissues that experience cyclic loading, opening up new avenues for exploring soft tissue homeostasis. 
\end{abstract}%

\section*{Introduction}

During an average lifetime, the human heart epithelium undergoes more than 2.5 billion contraction and expansion cycles \cite{Humphrey2002}, far surpassing the durability of most synthetic polymers. Epithelial tissues are two-dimensional sheets which line the internal structures and cavities of the body and are often monolayered, but can also be stratified or pseudo-stratified. They possess remarkable resilience to deformations and function as physical barriers. In dynamic mechanical environments, epithelia are challenged by intrinsic and extrinsic stresses, to which they respond over different length- and time-scales, from protein recruitment, to signalling and metabolic adaptation. The mechanical durability of epithelial monolayers raises important questions about the molecular and biophysical mechanisms underlying their resilience and is therefore an emerging research focus within mechanobiology. Here, we review the homeostatic and remodelling processes that preserve epithelial tissue integrity, focussing on the intrinsic timescales of epithelial remodelling and repair mechanisms. We propose that the integrity of epithelial tissues that undergo cyclic deformation requires a subtle balance between the timescales of damage and self-healing mechanisms and the period of the strain cycles.


\section{Epithelia commonly experience cyclic loading}
Cyclic loading is ubiquitous in normal physiology and manifests as periodic stretch and relaxation cycles in many organs. For instance, the heart undergoes oscillatory stress during contraction cycles and the intestinal lumen undergoes rhythmic deformation during peristalsis. Similarly, the bladder and lungs undergo periodic filling and voiding during physiological fluid flow. Although cyclic loading is essential for barrier function, excessive or abnormal loading can lead to tissue fatigue, micro-damage, and rupture. While some damage is a direct consequence of abnormal loading, other damage is indirect, arising through perturbation of processes normally regulated by cyclic loading. Indeed, cyclic deformation during normal breathing (25\% strain) regulates molecular processes involved in cytoskeletal regulation and barrier function \cite{Geiger2009}, while pathological strains ($>$37\% in mechanical ventilation) can lead to altered junctional protein expression and paracellular gap formation, compromising barrier function \cite{Davidovich2013}. Excessive or dysregulated cyclic stretch, characterised by perturbed stretch amplitude and frequency, has also been linked to pathologies such as heart valve prolapse, acute lung injury, and inflammatory bowel disease \cite{ el2021valve,kollisch2014mechanical,Sarna2010}. Despite its importance, the relationship between specific cyclic loading regimes and tissue integrity has not been systematically explored.

\begin{figure}
\centering    
\includegraphics[width=\linewidth]{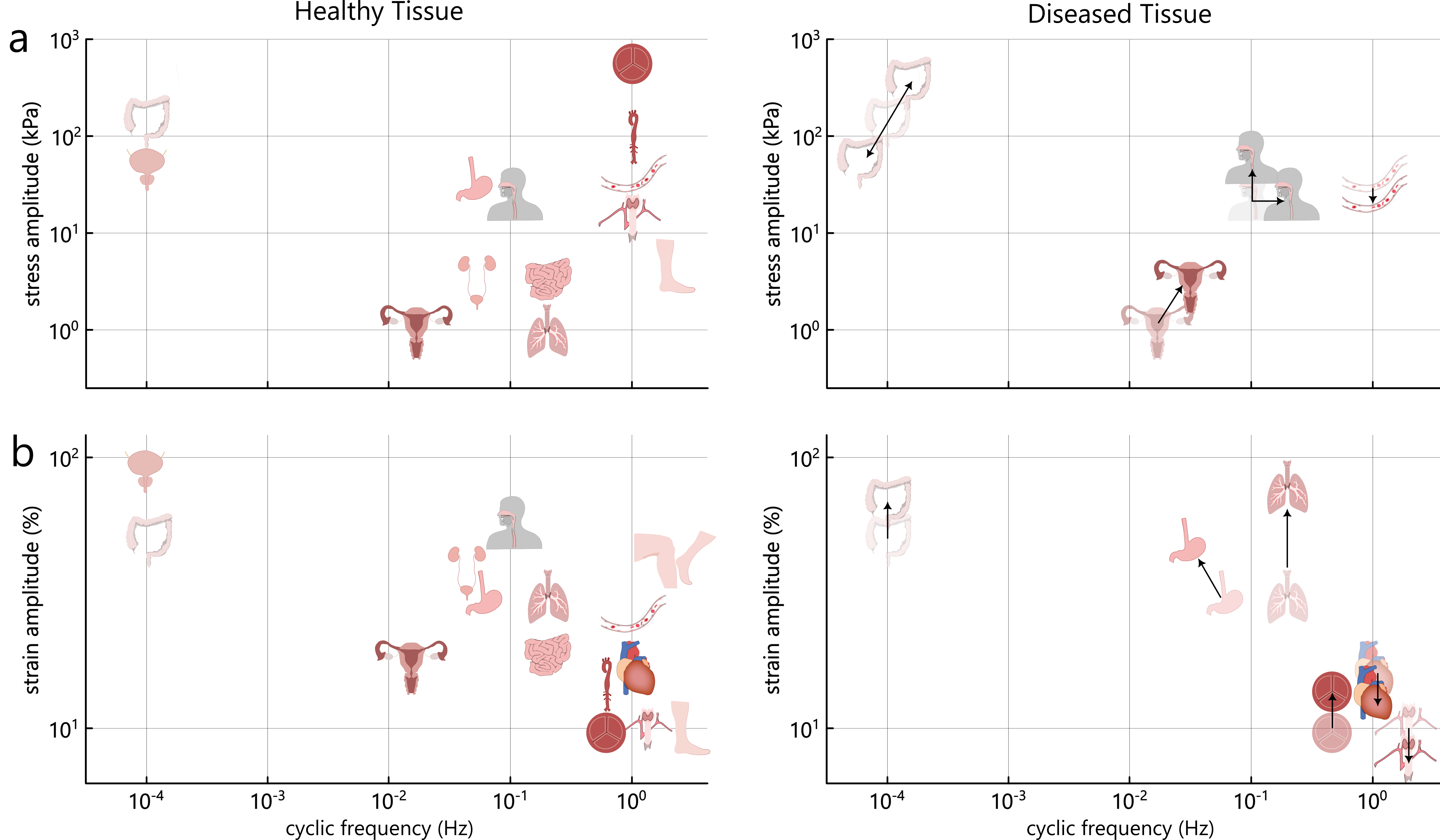}
\caption{a) Stress (kPa) and b) strain (\%) amplitudes over frequency (Hz) for healthy (left column) and diseased (right column) epithelial tissues within different organs. Supporting data with exact values and references can be found in Table \ref{tab:cycling}. Some icons have been adapted from NIAID NIH BIOART Source (bioart.niaid.nih.gov/bioart).}
\label{fig:cyclingparameters}
\end{figure}

To provide an overview of common physiological and pathological cyclic deformations, we summarise data from studies which report cyclic loading regimes either in terms of stress amplitude vs frequency or strain amplitude vs frequency (Figure \ref{fig:cyclingparameters}, Table \ref{tab:cycling}). Stress measurements in epithelial tissues are often challenging to interpret due to the difficulty in isolating the contribution of the epithelium from that of coupled tissues, such as the stroma (comprising extracellular matrix and stromal cells) or underlying muscle layers. As a result, stress experienced by the epithelium alone is rarely reported. In contrast, strain offers a more reliable metric, as all coupled layers of the tissue deform cohesively to the same extent under mechanical loading. Cyclic strain characterisation is therefore a valuable metric for studying strain-controlled remodelling and self-healing in epithelia.

\newgeometry{left=1.5cm, right=1.5cm, top=1.5cm} 

\begin{table}[]
    \centering
    \begin{tabular}{llllc}
    \toprule
     Tissue &  Stress Amplitude & Frequency &  Strain Amplitude & Icon \\
    \midrule
      lung & 1kPa \cite{faisy2011airway} &   0.2Hz \cite{scott2020monitoring} & \makecell[l]{30\%\cite{grune2019alveolar}\\ \textcolor{gray}{[MV 84\% \cite{kollisch2014mechanical}]}} & \includegraphics[height=1cm]{ 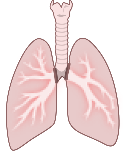} \\
      aorta & 100kPa \cite{Homan2023,west1991stress} &  1Hz \cite{persson2024reference} & 10-15\% \cite{Humphrey2002}& \includegraphics[height=1.0cm]{ 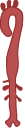} \\
    heart & - & 1Hz \cite{persson2024reference} & \makecell[l]{16\% \cite{howard2016left} \\ \tg{[ischemia 12\% \cite{howard2016left}]} }& \includegraphics[height=1.0cm]{ 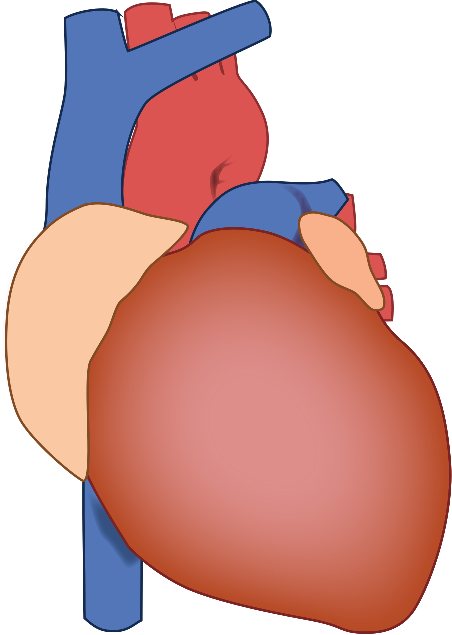} \\
      carotid & 15kPa \cite{Soleimani2021} &   1Hz \cite{persson2024reference} & \makecell[l]{10\% \cite{forsblad2021biomechanical} \\ \tg{[AS 8\%\cite{forsblad2021biomechanical}]}}& \includegraphics[height=1cm]{ 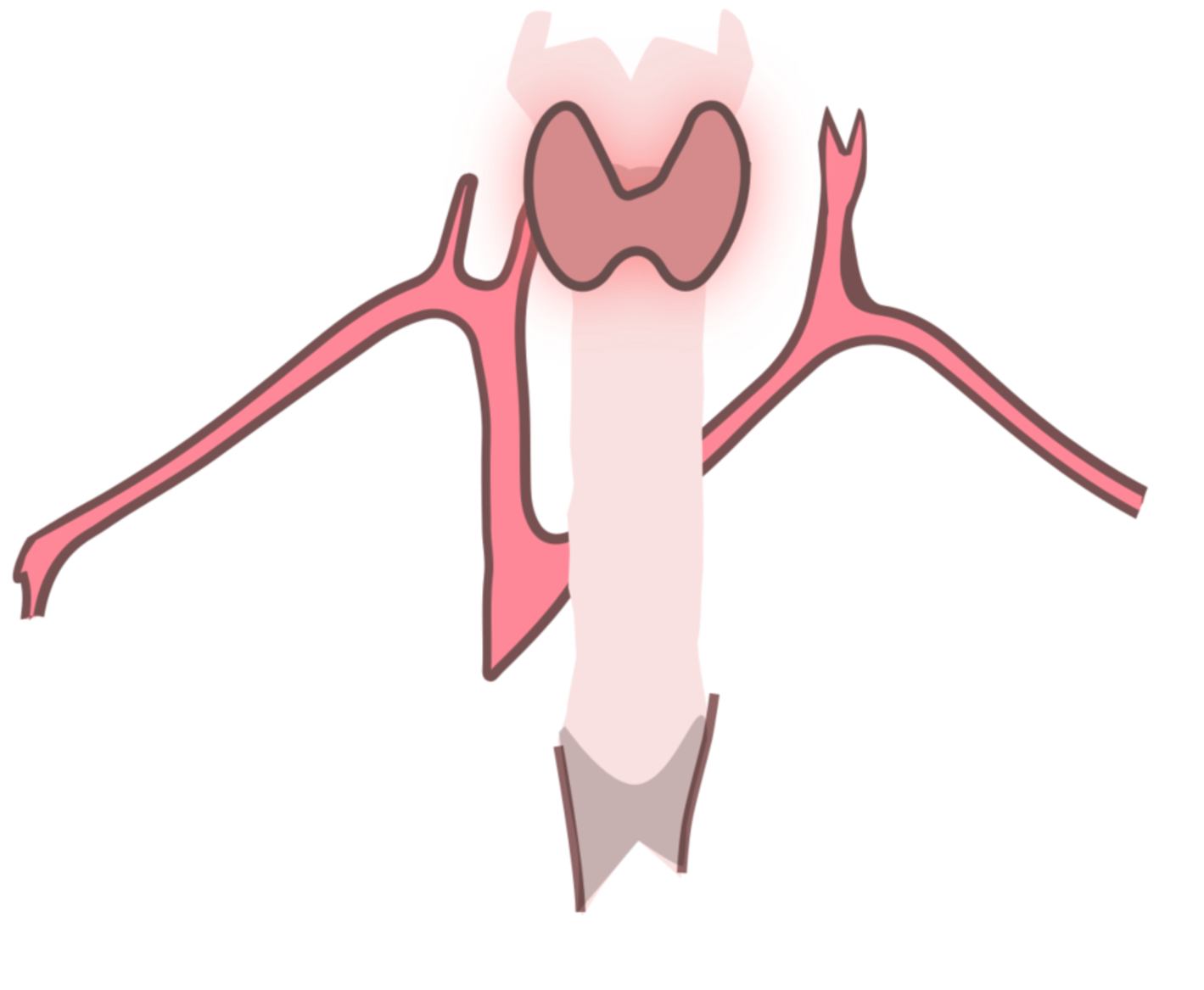} \\
      capillaries & \makecell[l]{35kPa \cite{neligan2024fluid} \\ \tg{[thrombosis 29kPa \cite{seem1990transcapillary}]} }&   1Hz \cite{persson2024reference} & 25\% \cite{HENQUELL1976259}& \includegraphics[height=1cm]{ 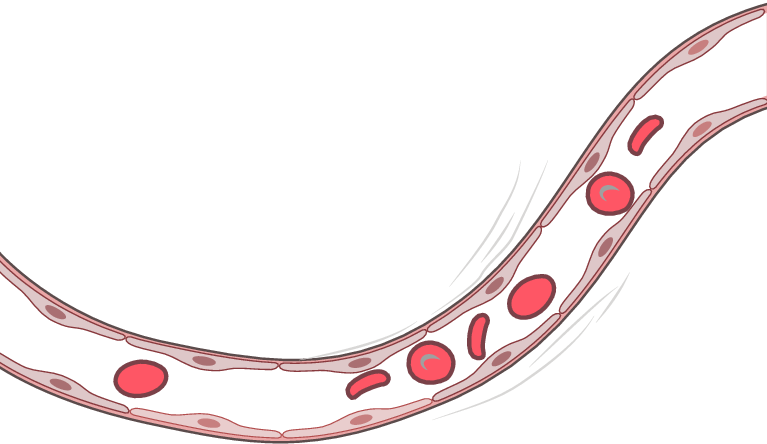} \\
      heart valve & 
    \makecell[l]{600kPa AV,\\ 100kPa PV \cite{parfeev1983mechanical}}
 &    1Hz \cite{persson2024reference}  & \makecell[l]{10\%  \cite{Dalgliesh2019} \\ \textcolor{gray}{[MVP 12\% \cite{el2021valve}]}} & \includegraphics[height=1cm]{ 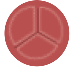} \\
      colon  & \makecell[l]{50kPa RPC \cite{Sarna2010},\\ 130kPa GMC \cite{Sarna2010}\\ \textcolor{gray}{[IBS-D GMC 250kPa \cite{Sarna2010}]} \\ \textcolor{gray}{[IBS-C GMC $<$65kPa \cite{Sarna2010}]}} & \makecell[l]{0.02Hz slow RPC \cite{Sarna2010} \\ 0.1Hz fast RPC \cite{Sarna2010} \\ $10^{-4}$Hz GMC \cite{Sarna2010} \\ \tg{[IBS-D GMC $>$$10^{-4}$Hz \cite{Sarna2010}]} \\ \tg{[IBS-C GMC $<$$5\cdot 10^{-5}$Hz \cite{Sarna2010}]}} &                                              \makecell[l]{50\% \cite{jaffe2015large} \\ \tg{[obstruction } \\ \tg{\ $\ \ >$50\% \cite{jaffe2015large}]}} & \includegraphics[height=1cm]{ 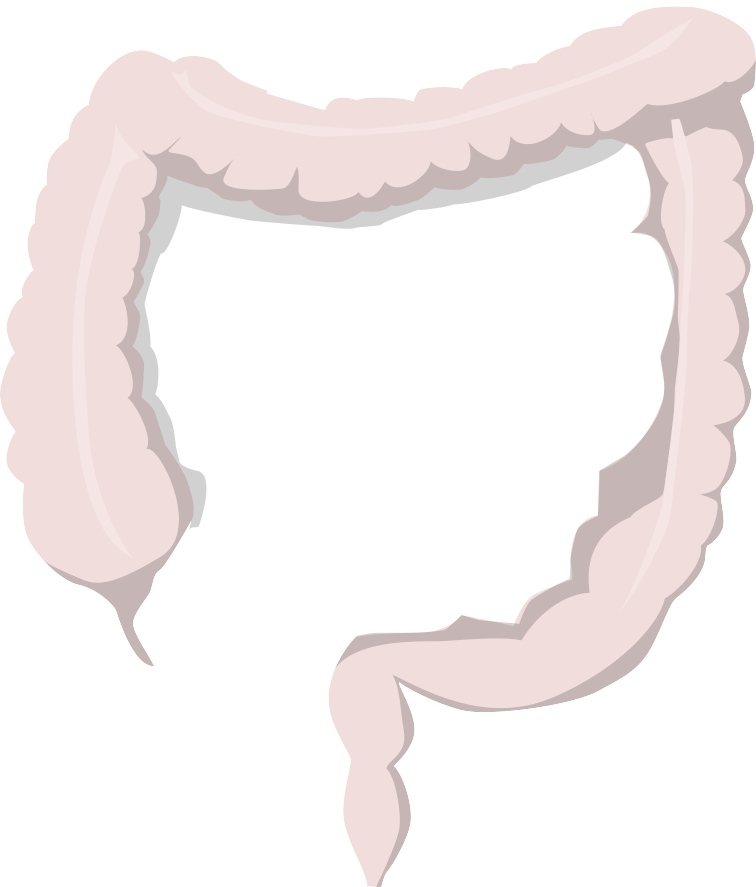} \\
      skin - walking & 5.5kPa \cite{Joodaki2018} & 2Hz \cite{rowlands2007influence} & 10\% \cite{Figueiredo2019CorrelationOL} & \includegraphics[height=1cm]{ 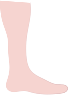} \\
      skin - running & - & 3Hz \cite{rowlands2007influence} & 43\% \cite{Figueiredo2019CorrelationOL}& \includegraphics[height=1cm]{ 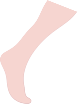} \\
      skin - knee flexion & - & 2Hz \cite{rowlands2007influence} & 45\% \cite{6294435}& \includegraphics[height=1cm]{ 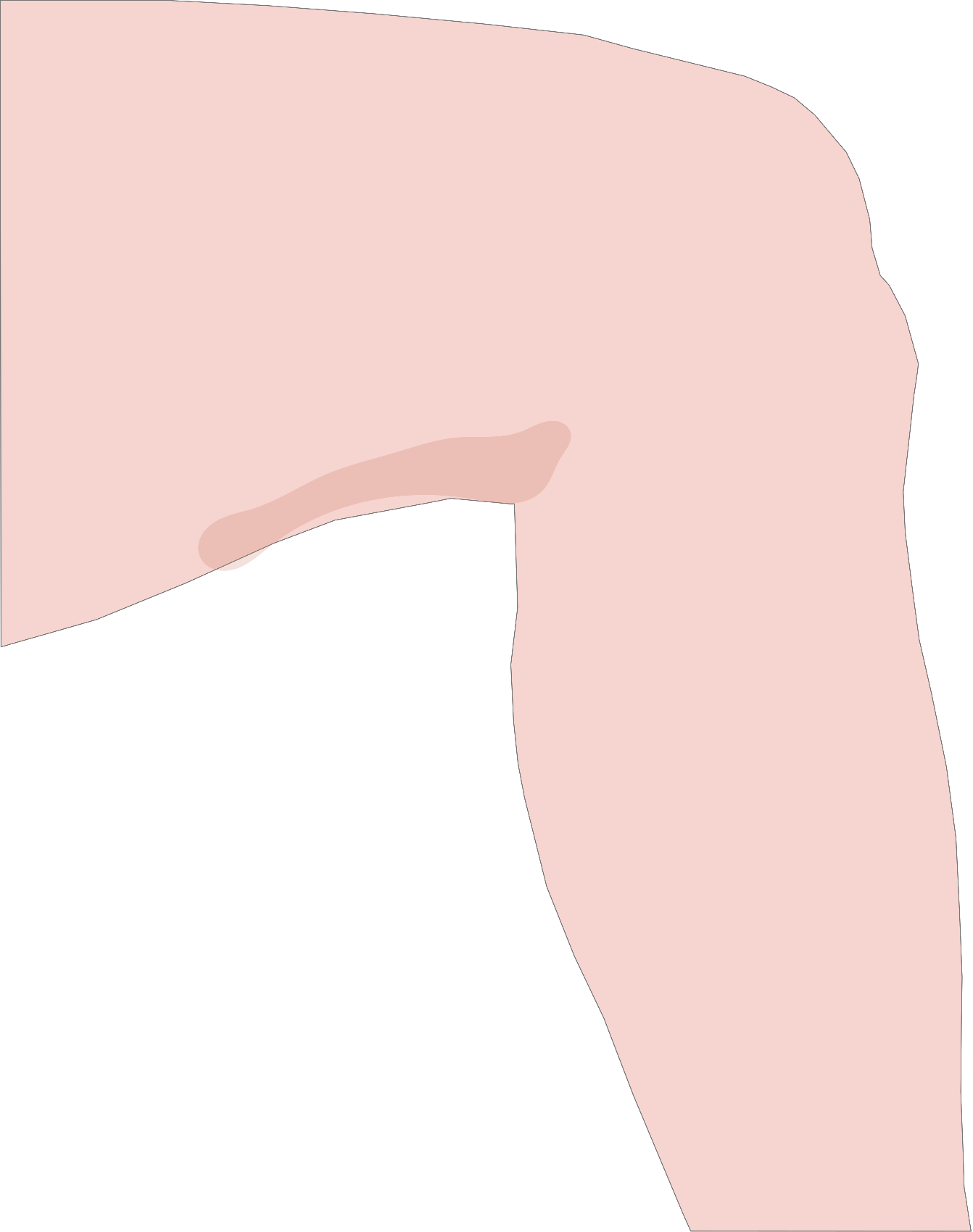} \\
      stomach &   30kPa \cite{indireshkumar2000relative}  & \makecell[l]{0.05Hz \cite{Ebara2023} \\ \tg{[FP 0.03Hz \cite{kayar2016gastric}]} } & \makecell[l]{30\% \cite{bharucha2011gastric} \\ \tg{[FP 48\%\cite{bharucha2011gastric}]}} & \includegraphics[height=1cm]{ 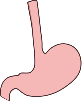} \\
      small intestines &  2.4kPa \cite{schmidt1996ambulatory} & 0.2Hz \cite{Zhang2006} & 20\% \cite{Zhang2006} & \includegraphics[height=1cm]{ 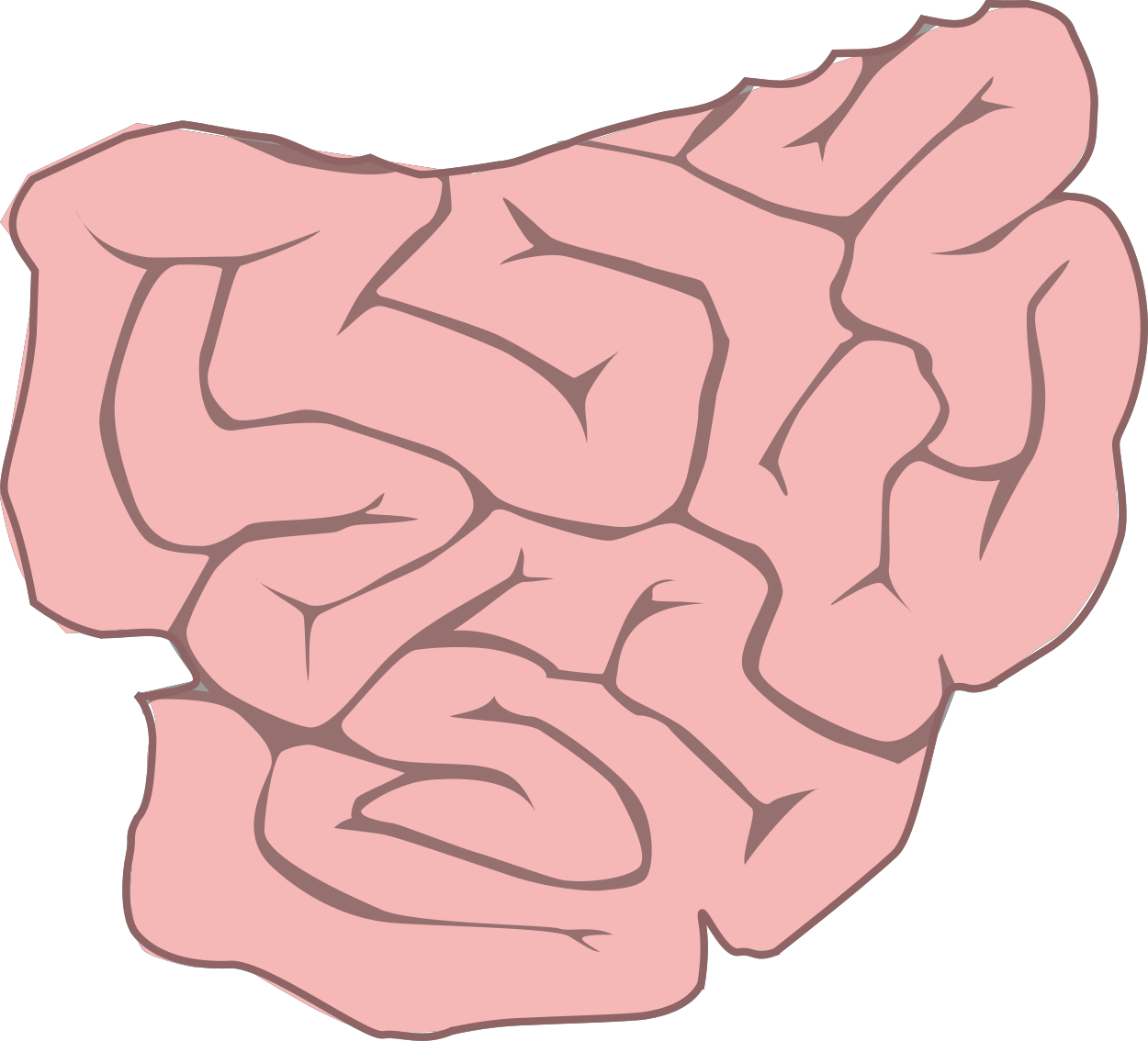}\\
      bladder & 51kPa \cite{PAREKH20101708} & $10^{-4}$Hz \cite{wyman2022urination} & 90\% \cite{PAREKH20101708}& \includegraphics[height=1cm]{ 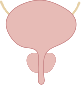} \\
      uterus & \makecell[l]{1kPa  \cite{BULLETTI20021156} \\ \tg{[endo. 3kPa \cite{BULLETTI20021156}]}} &  \makecell[l]{0.015Hz \cite{BULLETTI20021156}\\ \tg{[endo. 0.03Hz \cite{BULLETTI20021156}]}} & 16\% \cite{eytan2001characteristics} & \includegraphics[height=1cm]{ 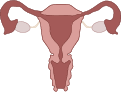} \\
      ureter  &  3kPa \cite{roshani2002intraluminal} & 0.05Hz \cite{hickling2017anatomy} &  35\% \cite{keni2024analyzing} & \includegraphics[height=1cm]{ 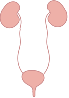} \\
      esophageal & \makecell[l]{  25kPa \cite{10.1001/jama.1983.03340190057032} \\ \tg{[dysphagia 60kPa \cite{10.1001/jama.1983.03340190057032}]}} & \makecell[l]{0.1Hz \cite{carlson2020rhythm} \\ \tg{[achalasia 0.2Hz \cite{carlson2020rhythm}]}} & 50\% \cite{leibbrandt2016characterization, lee2012esophageal} & \includegraphics[height=1cm]{ 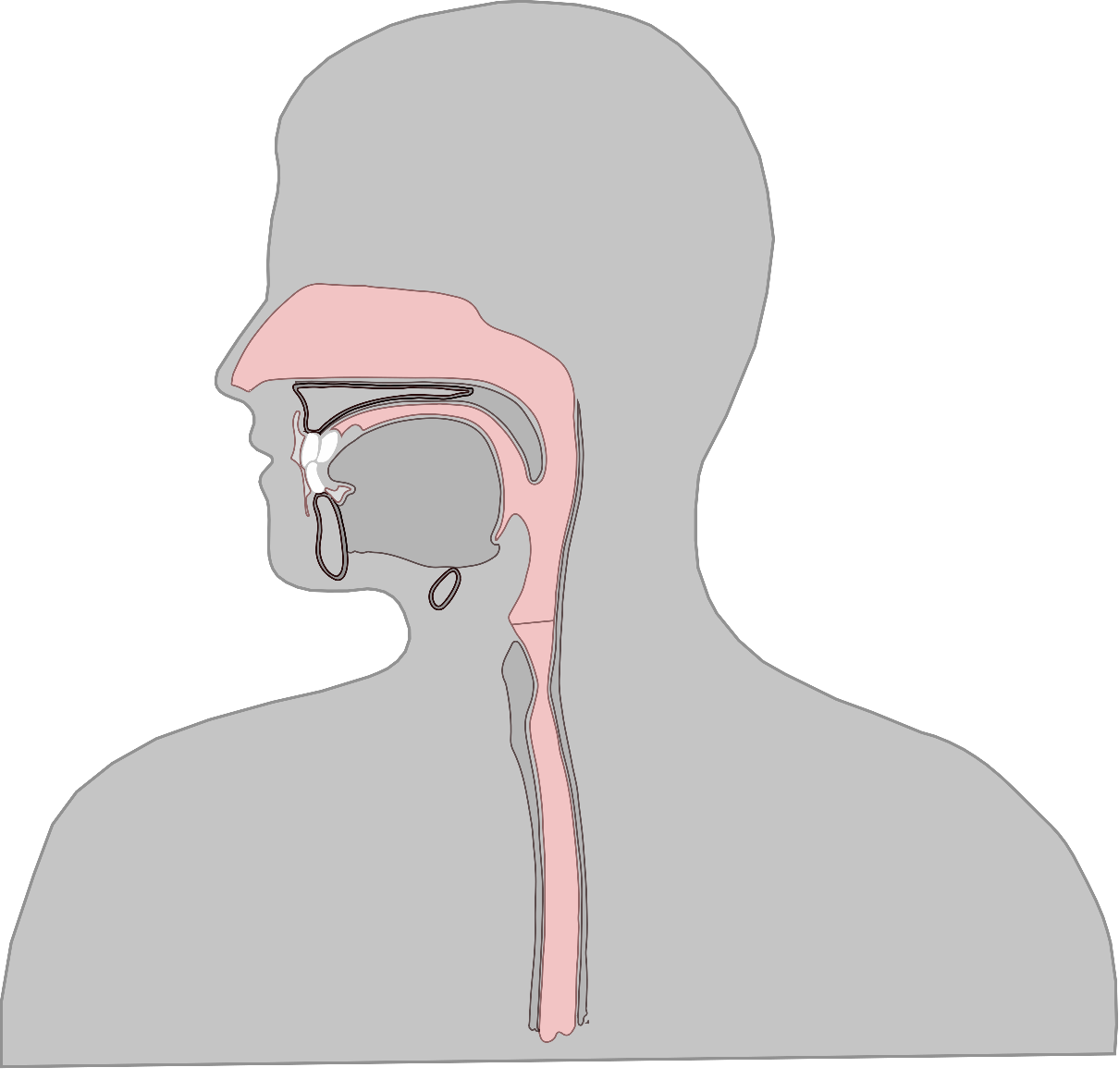} \\
\bottomrule
    \end{tabular}
    \caption[Stress and strain cyclic loading parameters]{Stress and strain cyclic loading parameters for organs lined by epithelia in health and disease. \\\hspace{\textwidth} Nomenclature: MV=Mechanical ventilation, AS=Ankylosing Spondylitis, AV=aortic valve, PV=pulmonary valve, MVP=Mitral valve prolapse, 
    RPC=Rhythmic phasic contractions, GMC=giant migrating contractions, FP=functional dyspepsia, endo.=endometriosis.}
    \label{tab:cycling}
\end{table}

\restoregeometry

\section{Mechanical response of epithelia to strain}

\begin{figure}
    \centering
    \includegraphics[width=1\linewidth]{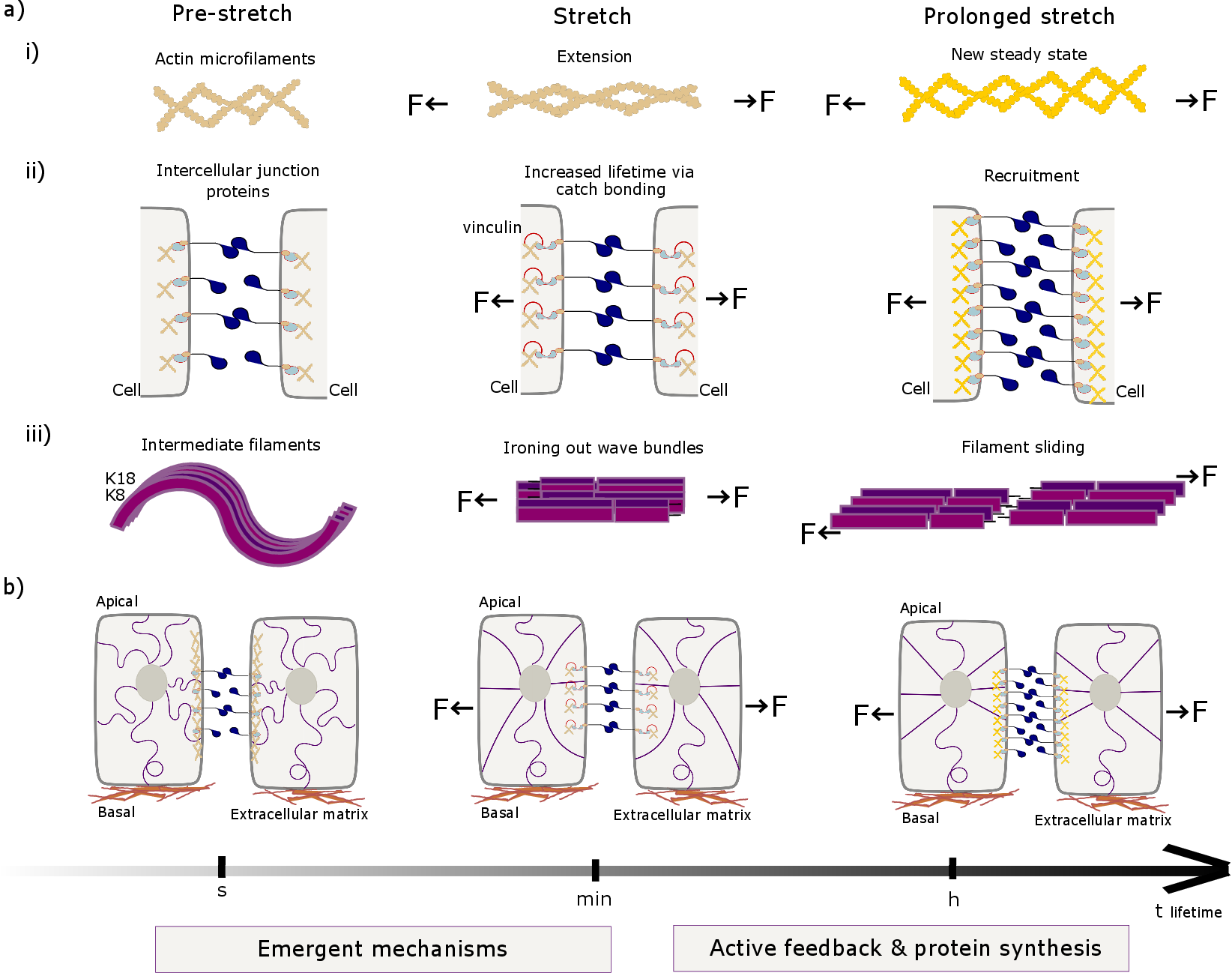}
    \caption{ \textbf{Response to deformation across length- and time-scales in epithelial monolayers.} For each row, the left column represents the organisation before stretch, the middle column immediately after application of stretch, and the right column after prolonged stretch. a) Mechanisms of plasticity in cellular and molecular components in response to a step loading in strain. i) Actin filaments become stretched but may adopt a long-term new steady state with an increased number of actin filaments, ii) Intercellular junction proteins, such as alpha-catenin and E-cadherin, behave as catch bonds, becoming stabilised upon loading. Alpha-catenin unfolds to recruit vinculin and redistribute the applied load to the cadherin-catenin force chain. At longer time-scales, protein recruitment will return cadherin-catenin complexes to their homeostatic load, iii) Keratin filament bundles become loaded in response to stretch but dissipate stress by inter-filament siding in prolonged loading. b) Time-scales associated with emergent mechanisms and active feedback for cytoskeletal reorganisation, turnover, and protein synthesis. $\rm t_{lifetime}$ refers to the functional stabilization time-scale for the mechanisms of plasticity}
    \label{fig:scales}
\end{figure}

In response to deformation, the stress that epithelia experience depends on their rheology, which has been shown to be nonlinear and highly dependent on the magnitude and rate of strain. All levels from single molecular crosslinkers \cite{Lee2009}, to individual filaments \cite{Block2017} and filament networks \cite{Kechagia2024} contribute unique time-dependent behaviours (Figure \ref{fig:scales}). Non-linear behaviours such as strain stiffening may represent an emergent adaptation of the tissue to resist sudden and transient increases in loading, such as shocks. Notably, epithelia can stiffen under strain, depending not only on the magnitude, but also the rate of deformation \cite{Duque2024}**. If dysregulated loading is sustained, cells within the epithelium adapt by regulating both the cytoskeleton and junctional proteins, which act together to resist mechanical stresses and preserve barrier function. Adaptation can take place through post-translational modifications, protein recruitment (second to minute time-scales), or transcriptional regulation (longer time-scales) \cite{Pinheiro2018,Dupont2022} (reviewed in \cite{Campas2024}**). Depending on the time-scale of cyclic loading distinct biophysical phenomena may contribute to damage and repair. 

\subsection{The dynamic actin cytoskeleton makes tissues resilient}
The actin cytoskeleton, connected across cells via adherens junctions (AJs), forms a supracellular network that plays a key role in tissue mechanics. In AJs, E-cadherin binds to neighbouring cells through its extracellular domain and interfaces with actin through binding its intracellular domain to beta- and alpha-catenin. Both the actin cytoskeleton and AJs respond to and are regulated by mechanical tension. Actin networks exhibit a linear tension-strain response under small deformations and are regulated by mechanical forces \cite{Wioland2019}. At longer timescales, actin networks behave as viscous fluids due to turnover of crosslinkers and actin filaments ($\leq$1min, \cite{Fritzsche2013}) (Figure \ref{fig:scales}ai), leading to stress relaxation in epithelial tissues \cite{Khalilgharibi2019}. In immediate response to load, several AJ proteins exhibit catch-bonding characteristics, where molecular bonds become more stable under a moderate load (Figure \ref{fig:scales}aii). E-cadherin trans-dimers, which bind adjacent cells to one another, are most stable under loads of around 30 pN \cite{Rakshit2012}. The bond between actin filaments and alpha-catenin is stabilized 20-fold under 10 pN molecular load \cite{Buckley2014}. Similar behaviour also emerges in other intercellular adhesion proteins associated with the actin cytoskeleton: PDZ-domains from nectin-1 and JAM-A were recently shown to form catch bonds with afadin \cite{Vachharajani2023}*. Interestingly, recent structural evidence suggests that nectin rather than E-cadherin may be the primary junctional protein involved in force transmission through the supracellular actin cytoskeleton in mature gut epithelia \cite{Mangeol2024}**. Catch-bonding enables cells to rapidly strengthen intercellular junctions by extending bond lifetimes and forming new links, buffering transient tension. Vinculin recruitment reinforces cadherin-catenin complexes for added stability. Under prolonged loading, additional complexes are recruited, re-distributing the load to reduce each bond's lifetime and restore the proportion of bound links to homeostatic levels (Figure \ref{fig:scales}aii).


\subsection{Intermediate filaments act as shock absorbers and provide mechanical memory}
 

Intermediate filaments (IFs) and the desmosomes which connect them between neighbouring cells form a second supracellular network within tissues. Although evidence is conflicting on whether desmosomal proteins bear load while the epithelium is at rest \cite{Price2018, Baddam2018}, the emergent properties of the keratin network connected by desmosomes dominate the mechanical response at large deformations \cite{Duque2024}**. At rest, bundles of keratin filaments appear wavy. When the tissue is stretched they gradually straighten and become progressively load bearing under large deformation \cite{Harris2012}, leading to a multi-stage strain stiffening response (Figure \ref{fig:scales}aiii). To capture this, recent modelling studies incorporate nonlinear components for the rate-dependent strain-stiffening of supracellular cytoskeletal networks, such as a `slack' representing the waviness or `degree of entanglement' of keratin bundles \cite{Duque2024}, acting as an emergent topological safety net \cite{Pensalfini2023}.
Under sustained stretch, individual filaments can slide past one another \cite{Lorenz2023}**, leading to intra-bundle flow and energy dissipation over tens of minutes. This mechanism may confer a long-term mechanical memory of the epithelium's deformation history. Mechanotransductory responses associated with keratin IFs are less well studied. Some pathways increase interfacing of the keratin network with the actin cytoskeleton via the cytolinker plectin in response to tension \cite{Prechova2022}\cite{Ruiz2024}**. Tensin 4 is recruited to keratin fibres subjected to tension and remains localised for minutes after deformation release \cite{Cheah2019}, suggesting a role in adapting to cyclical loads with periods shorter than minutes.
At longer timescales, keratins may indirectly regulate transcriptional responses by shielding the nucleus from deformation \cite{Laly2021}. Adaptations of desmosomal junctions to stress may also contribute to mechanical memory, but the extent and timescale of mechanoresponsive desmosome remodelling remain less well-characterized \cite{broussard2020scaling}.
 
\subsection{Non-physiological strain leads to damage accumulation in living materials}
Damage in tissues can result either from a single catastrophic high-amplitude deformation 
or from accumulation of defects due to repeated cycles of supraphysiological strain. 
We can apply the concept of fatigue from material engineering, where controlled cyclic loading is used to probe a material's endurance limits, energy dissipation, and failure mechanisms. In classic engineering materials, fatigue leads to cumulative damage - a progressive and irreversible process where repeated stress cycles gradually degrade the material's structure \cite{Lee2005}. Even when individual cycles do not cause immediate macroscopic damage, microscopic lesions accumulate incrementally until failure. Living tissues, by contrast, possess repair mechanisms that will be discussed in Section \ref{repair}.

In soft tissues, damage accumulation can either directly result from rupture of intercellular bonds and ECM adhesions, or indirectly arise from active responses of cells to strain. Cyclic stretch testing in mouse alveolar epithelial layers reveals that intercellular gaps form under hyperinflation (80\% change in lung capacity), leading to a loss of epithelial barrier integrity. Intercellular junctions are destabilised due to downregulation of p120-catenin which induces paracellular gap formation \cite{Wang2011}. Similarly, large amplitude uniaxial cyclic stretch (70\%) of the oesophageal mucosa demonstrated hysteresis, permanent deformation, stress softening, and occasionally rupture \cite{Durcan2022}. Finally, mutations in keratin 5 or keratin 14 in epidermolysis bullosa simplex, have been linked to weakening of the epithelial sheet integrity upon shear stress \cite{Homberg2015}.


Geometry is emerging as a critical factor in destabilising epithelia: the orientation of forces acting on junctional complexes is likely a major determinant in how junctions respond to load. For instance, tensile stress on AJs increases both E-cadherin and vinculin recruitment, whereas shear stress leads to loss of E-cadherin from AJs in the Drosophila germband \cite{Kale2018}\cite{Nishizawa2023}**. Such geometry-dependent stability of junctions may explain why disease-related changes in junctional actin organisation lead to tissue fragility. For instance, pathological expression of the actin-bundling protein fascin-1 modifies the geometry of AJs by changing the curvature and alignment of junctional interfaces. This reduces junction stability and enhances cellular protrusive activity, leading to elongated shapes and polarization, which can induce cancer cell migration and invasion \cite{Esmaeilniakooshkghazi2023}*. 

Further detailed studies of the structures of intercellular contacts in epithelia are essential for improved understanding of how different cycling regimes affect epithelial stability. Such studies are becoming increasingly feasible thanks to recent advancements in the imaging of intercellular junctions (reviewed in \cite{Janssen2024}*). 

\section{Repair mechanisms in epithelial tissues} \label{repair}

Living materials are particularly intriguing when considering fatigue: damage can often be repaired through healing and remodelling processes. In this section, we examine mechanisms of repair at the molecular- and cellular-scale. 


\subsection{Re-binding of intercellular linkers}

Repair at the molecular scale can either take the form of (i) gap or micro-lesion closure (when two surfaces have fully separated) or (ii) return to homeostatic intercellular bond density, where no significant intracellular gap has formed but the number of links connecting two cells has decreased due to force application. 

\begin{figure}
    \centering
    \includegraphics[width=1.0\linewidth]{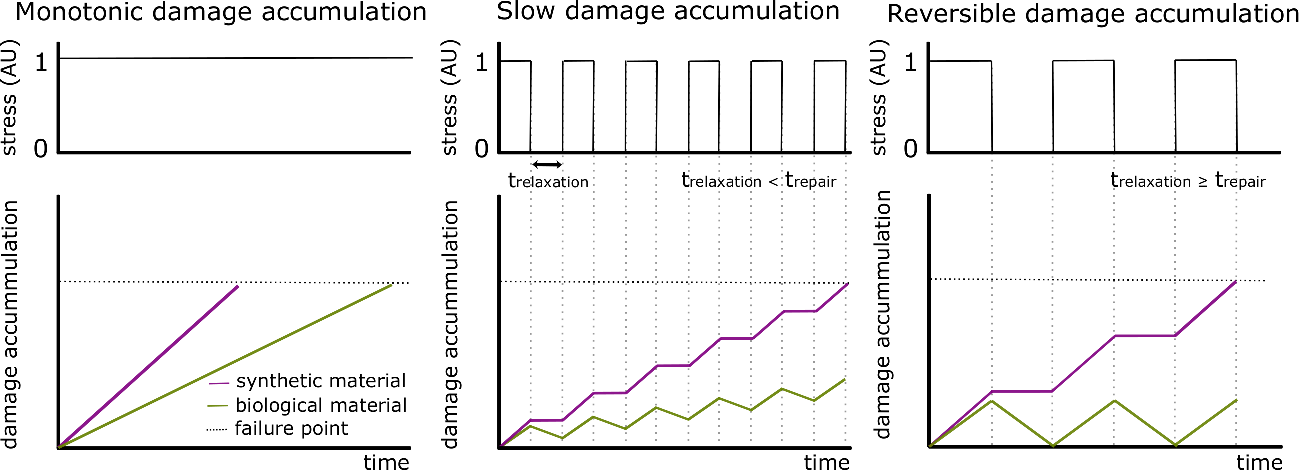}
    \caption{Comparison of damage accumulation signatures for synthetic and living materials under fixed and cyclic loads (conceived in the context of intercellular junction dynamics but extends to other repair mechanisms). Under constant stress, damage accumulation in synthetic materials is a monotonic function (purple) until failure (black, dashed), similarly to living materials (green). Under cyclic stress, synthetic materials accumulate damage, but living materials may repair part or all of the accumulated damage. We speculate that the damage and repair mechanisms depend on the relationship between the relaxation portion of the cycle ($\rm  t_{relaxation}$) and the time-scale of the repair mechanism ($\rm  t_{repair}$). The linear repair functional form is a naive representation and may vary depending on the material, repair mechanism and imposed loading waveform, which haven't been comprehensively explored.}
    \label{fig:fatiguerepair}
\end{figure}

A well-established model to study the biophysics of intercellular adhesion was developed by Bell  \cite{Bell1978}. A cell junction is defined as two surfaces connected by reversible links which can be bound or unbound. Each link has a constant binding probability $\rm k_{on}$, and an unbinding probability that depends on force, $\rm k_{off} = \rm k_{off,0}e^{\frac{\rm f}{\rm f_0}}$. Rupture occurs when all linkers connecting the two surfaces are unbound. This model enables the exploration of critical loading conditions — such as strain ramps applied at varying rates — that result in junction failure. Previous studies have shown that the outcomes depend on the interplay between the binding and unbinding rates, the rate at which the load is applied, and the magnitude of the load itself \cite{Erdmann2004,Duque2024,Mulla2018}. Above a critical constant load, links are rapidly lost and failure ensues. However, the impact of time-varying loads has not been comprehensively explored, especially when loads only transiently surpass the critical load. An interesting illustration of such a transient load comes from the \textit{Drosophila} amnioserosa, which displays pulsative forces with a ~200$\,$s periodicity. When embryos are extrinsically loaded \cite{Sumi2018} or when myosin becomes over activated \cite{Duque2016}, gaps appear between cell surfaces. Myosin brings delaminated surfaces back together within 200-400$\,$s \cite{Sumi2018}. Figure \ref{fig:fatiguerepair} illustrates how periods of low tension may allow for repair and self-healing in living materials. Thus, repair mechanisms may allow tissues to reach high tensions non-destructively for a limited period of time in the context of cyclic loading. 

A key assumption of Bell's model is that force is distributed globally across all bound links. Subsequent extensions have implemented local load sharing to study crack initiation and critical length for junction failure \cite{Mulla2018}. While most models assume a rigid cell membrane, Li et al. accounted for viscoelastic cellular deformation \cite{Li2016}. Their findings suggest that viscosity can favour repair by prolonging the time it takes to separate membranes, thereby enhancing the likelihood of ligand-receptor pair rebinding. Once surfaces separate beyond the reach of intercellular linkers, cellular protrusions are needed to reconnect them. Consistent with this idea, recent work suggests that protrusive forces constantly act at junctions to maintain close membrane contact \cite{Senju2023}*. 

In cells, E-cadherins and desmosomal cadherins both assemble in distinct clusters whose size is force regulated \cite{Troyanovsky2023}. Forces below 10$\,$pN are predicted to create large numbers of small clusters (less than 5 E-cadherins each), while high forces result in a higher probability of forming few but large clusters \cite{Chen2021}*. How this regulation takes place and why adhesions are organised in clusters remains unclear. One intriguing suggestion is that nanoclusters are the result of interdigitated microspikes extended by each cell to increase the area of intercellular contact. From a biophysical perspective, arranging links into clusters may enable junctions to be more adaptable and repair more efficiently. This raises further questions such as how forces are distributed among and within clusters, and how this may contribute to junction remodelling. 

\subsection{Wound healing through tissue fluidisation}

Once a cellular or multicellular wound has appeared, different mechanisms are necessary to close the gap. These involve lamellipodial protrusion, to invade the gap, or the creation of a multicellular actin purse string that brings cell membranes into contact to regenerate new junctions. Closure of multicellular gaps necessitates reorganisation of the tissue involving cell intercalations. For example, in response to a multicellular wound, the \textit{Drosophilla} wing disk epithelium first undergoes an initial fast closure phase ($\sim $20$\,$min) attributed to myosin II-mediated purse string contraction which closes the gap at a rate of $\sim$$15\, \mu$m$^2$/min \cite{Kobb2017}, followed by a slow phase of edge cell intercalations lasting up to four hours that necessitates unjamming of the epithelium\cite{Tetley2019}. Additionally, neighbour exchange in epithelial cells is reminiscent of a brittle to ductile transition as a result of topological defects which accelerates wound repair. During convergent extension, the  Drosophila germ band undergoes 50\% tissue strain and cell intercalation occurs over 25-30$\,$min \cite{Simoes2014}. During this time, the original junction shrinks to a four-way contact point within 10$\,$min and it takes another 10$\,$min to build a new junction, completing neighbour exchange. How cells preserve tissue integrity during this process remains an area of active investigation. 

Intriguingly, cyclic loading may favour intercalation by transiently decreasing the load on junctions. Indeed, experiments and modelling indicate that local heterogeneity in tension drives neighbour exchanges \cite{Curran2017}. This suggests that physiological cyclic loading may promote remodelling and neighbour exchanges because of fluctuating tension at intercellular contacts. In support of this, unidirectional cyclic stretch was found to increase the frequency of cell rearrangements and induce elongation of epithelial cell colonies in the direction of stretch \cite{Comelles2021}. Mathematical modelling suggested that alterations in cell shape along with directional expansion of cell-cell interfaces can induce cell intercalation through mechanisms dependent on myosin II activity \cite{Lien2023}*.
Overall, by comparing the frequency and duration of external cyclic mechanical stimulation with the intrinsic timescales of cell migration, proliferation and intercalation, we can extract valuable insights about which tissue repair mechanisms may be at play in the different epithelial tissues subjected to cyclic loading.

\section{Conclusion}

To grasp the conditions required for the stability of epithelial tissues under physiological cyclic deformations, it is crucial to consider the rich variety of damage and repair mechanisms in epithelial tissues, and the broad range of timescales at which they operate. We might speculate in particular that any damage accumulated during the stretch phases would be transient and repaired in the relaxed phases of the cycles. 

Therefore, studying the impact of cyclic loading provides a powerful framework to dissect the interplay between damage accumulation and adaptive repair processes. 
Experimental and theoretical model systems provide a rich test-bed for these ideas. 
Refining our understanding of the timescales of damage, repair and remodelling mechanisms in epithelial tissues will be important to address questions about living materials' homeostasis and self-healing.

\section*{Acknowledgements}
E.P. was supported by the Engineering and Physical Sciences Research Council Centre for Doctoral Training in Sensor Technologies for a Healthy and Sustainable Future [EP/S023046/1], L.B. was supported by an EMBO Postdoctoral Fellowship (grant no. ALTF 890-2023), L.B. and G.C. were supported by an sLoLa grant from the British Biotechnology and Biological Sciences Research council (BBSRC, grant no. BB/V019015/1) to G.C. and A.B. was supported by the European Research Executive Agency via the MSCA Postdoctoral Fellowships (BIOFRAC, Project number: 10110537).

\section*{Disclosure}
The authors declare no conflict of interest.




\end{document}